# Anomalous Hall effect in antiferromagnetic RGaGe (R = Nd, Gd) single crystals

Weizheng Cao[1#], Yuze Xu[1#], Yi Liao[1], Cuiying Pei[1], Juefei Wu[1], Qi Wang[1,2*], Yanpeng Qi[1,2,3*]

1. School of Physical Science and Technology, ShanghaiTech University, Shanghai 201210, China
2. ShanghaiTech Laboratory for Topological Physics, ShanghaiTech University, Shanghai 201210, China
3. Shanghai Key Laboratory of High-resolution Electron Microscopy, ShanghaiTech University, Shanghai 201210, China

    # These authors contributed to this work equally.
    * Correspondence should be addressed to Y.Q. (qiyp@shanghaitech.edu.cn) and Q.W. (wangqi5@shanghaitech.edu.cn)

## Abstract

Recently, the non-centrosymmetric Weyl semimetallic candidate family RTX (R = rare-earth element, T= poor metal, X = Si and Ge) has recently attracted significant attention due to its exotic quantum states and potential applications in quantum devices. In this study, our comprehensive investigations of high-quality NdGaGe and GdGaGe single crystals reveal distinct magnetic and electrical responses. Both compounds exhibit antiferromagnetic transitions with $T_N \sim$ 7.6 K and 22.4 K for NdGaGe and GdGaGe, respectively. NdGaGe exhibits strong magnetic anisotropy ($\chi_c/\chi_a \sim 70$). In contrast, GdGaGe displays weak magnetic anisotropic behavior ($\chi_c/\chi_a \sim 1$) with a distinctive spin-flop transition. Below $T_N$, NdGaGe shows significant negative magnetoresistance due to the reduced spin-disorder scattering arising from the field-induced spin alignment. GdGaGe exhibits more complex magnetoresistance behavior: positive values at low fields transitioning to negative values attributed to the reduced spin-flop scattering. Specially, NdGaGe demonstrates a large anomalous Hall conductance (AHC) of approximately 368 $\Omega^{-1}$ cm$^{-1}$, which is dominated by the intrinsic mechanism. These reveal the pivotal role of rare-earth elements in modulating the electronic structure, magnetic properties, and transport characteristics of the RGaGe system, thereby providing valuable insights for developing next-generation spintronic devices.

Strong electron correlations and electron-magnetic moment interactions in complex systems give rise to diverse physical phenomena, particularly in magnetic materials, among which the anomalous Hall effect (AHE) stands out as a prominent manifestation[1,2]. The AHE has garnered substantial interest across theoretical and experimental research, offering promising applications in energy-efficient electronics and spintronics[3-5]. The mechanisms of AHE can be categorized into intrinsic and extrinsic contributions[1]. On the one hand, the intrinsic Karplus–Luttinger (KL) mechanism originates from a nonvanishing Berry curvature, an intrinsic property of the Bloch wave function in systems where the time-reversal symmetry is broken[6-8]. Recently, a large intrinsic AHE due to the large Berry curvature effect arising from the nontrivial band topology has been reported in magnetic topological materials, such as $Fe_3Sn_2$, $Co_3Sn_2S_2$, $Mn_3Sn$, and $Co_2MnGa$[9-15]. On the other hand, the extrinsic mechanisms, including skew-scattering and side-jump scattering, arise from the asymmetric scattering of conduction electrons [16-18].

Recently, the tetragonal rare-earth-based compounds RAlX (R = La–Sm; X = Si and Ge), recognized as prominent representatives of magnetic Weyl semimetals with non-centrosymmetric structures, have garnered significant attentions[19-48]. These compounds demonstrate abundant magnetic ground states and remarkable electronic topology tunability by varying the rare-earth element R or metalloids X. In previous studies, ferromagnetic Weyl semimetal CeAlSi with in-plane noncollinear ferromagnetic order hosts a large anomalous Hall conductance AHC (~ 550 $\Omega^{-1}$ $cm^{-1}$)[22,25], arising from the Weyl fermions near the Fermi level ($E_F$) in $k$-space [45,46]. In ferromagnetic Weyl semimetal PrAlSi, while the Weyl points remain stable across the entire composition range when Ge is doped at Si sites, the $E_F$ position is effectively modulated. This modulation leads to the various magnitudes of AHE in PrAlGe$_{1-x}$Si$_x$ compounds and induces a transition from intrinsic to extrinsic mechanism[28,31]. NdAlSi shows a complex magnetic behavior, transitioning from a paramagnetic state to an incommensurate spin density wave state at 7.3 K, before establishing a commensurate up-down-down (UDD) ferrimagnetic order at 3.3 K[33-36]. NdAlGe displays a temperature-dependent transition from a down-up-up (DUU) ferrimagnetic state with Berry curvature-dominated AHC (~ 400 $\Omega^{-1}$ cm$^{-1}$) to a ferromagnetic state with significantly AHC (~ 1000 $\Omega^{-1}$ cm$^{-1}$) arises from chirality-driven AHE due to spin fluctuations and extrinsic disorders through skew-scattering[38,40].

In this study, we systematically investigate the magnetic and electrical transport properties of high-quality NdGaGe and GdGaGe single crystals, exploring the influence of different rare-earth elements on the physical characteristics in the RGaGe family. Both compounds undergo antiferromagnetic (AFM) transitions with $T_N$ ~ 7.6 K and 22.4 K for NdGaGe and GdGaGe, respectively. They demonstrate distinct antiferromagnetic behaviors, i.e. strong magnetic anisotropy for NdGaGe while a characteristic spin-flop transition within out-of-plane magnetic field for GdGaGe. Furthermore, it is found that NdGaGe exhibits a large AHC ( ~ 368 $\Omega^{-1}$ cm$^{-1}$) contributed by the intrinsic mechanism. This investigation reveals the modulation effect of rare-earth elements on electronic structure and their subsequent influence on the magnetic and transport properties of RGaGe family.

High-quality single crystals of NdGaGe and GdGaGe were synthesized by the Ga-In binary flux method. High-purity blocks of Nd/Gd, Ga, Ge and In were mixed in the molar ratio of 1:2:1:8 and loaded into an alumina crucible. All treatments were performed in an argon-filled glove box, and the crucible was sealed in a quartz tube under a vacuum. The quartz tube was heated up to 1050°C in 24 h with temperature holding for 12 h to ensure that the raw materials are melted. Subsequently, the temperature was slowly cooled down to 500 °C at a rate of 3 °C/h, and excessive flux was removed by a high-speed centrifuge.

The phase and quality examinations of both samples were performed on the Bruker AXS D8 Advance powder crystal X-ray diffractometer (XRD) with Cu K$_\alpha$ ($\lambda$ = 1.54178Å) at room temperature. The magnetization measurements were conducted on a magnetic property measurement system. Electrical transport measurements were performed on the physical property measurement system. The longitudinal and Hall resistivity were measured using a five-probe method. Furthermore, we extract the Hall resistivity by the equation $\rho_{yx}$ ($B$) = [$\rho_{yx}$(+$B$) - $\rho_{yx}$(-$B$)]/2 to remove the longitudinal resistivity contribution due to voltage probe misalignment.

The RGaGe compounds (R = Nd, Gd) crystallize in a tetragonal LaPtSi-type structure with space group $I4_1md$ (No. 109), isomorphic with LaAlSi[20]. The crystal structure consists of alternating layers of Nd/Gd, Ga, and Ge, with each layer containing only one element along the $c$ axis, as shown in Figure 1(a). To characterize the crystal structure, we conducted the powder XRD measurements. Figure 1(b) and (c) present the XRD patterns of the NdGaGe and GdGaGe single crystals, respectively. The powder XRD patterns were refined using the Rietveld method, confirming the phase purity of the samples. The refined lattice parameters of RGaGe compounds are summarized in Table S1. Figure 1(d) shows the temperature dependence of longitudinal resistivity $\rho_{xx}$ (T) for NdGaGe and GdGaGe single crystal. Both materials exhibit metallic behavior over the temperature range from 1.8 to 300 K. It should be noted that a dramatic drop was observed at around 7.6 K for NdGaGe and 22.4 K for GdGaGe, respectively, corresponding to the magnetic transition arising from the suppression of spin scattering [49,50].

The temperature-dependent magnetic susceptibilities $\chi$(T) with the zero-field-cooling (ZFC) and field-cooling (FC) modes for NdGaGe and GdGaGe single crystals were measured under an external field of $B$ = 0.1 T along $a$ and $c$ axis (Figs. 2(a) and 2(c)). The $\chi$(T) curves for NdGaGe and GdGaGe both display obvious cusps at low temperatures, corresponding to the antiferromagnetic transition at $T_N$ ~ 7.6 K and 22.4 K, respectively, coincide well with those values observed in $\rho_{xx}$(T) curves. For NdGaGe, the $\chi$(T) curves for $B$ // $c$ between ZFC and FC modes diverge significantly below $T_N$, possibly due to the spin reorientation. Moreover, it should be noted that the magnetic response in NdGaGe exhibits strong magnetic anisotropy, with $\chi_c$ approximately 70 times larger than $\chi_a$ at low temperature (Figs. S1), attributed to the single-ion crystal electric field (CEF) effects[51-53]. By contrast, in GdGaGe, the $\chi$(T) curves show a very weak magnetocrystalline anisotropy ($\chi_c$ /$\chi_a$ = 1.03), comparable to that of GdAuGe[54,55]. In the high-temperature region, the $\chi$(T) curves for both NdGaGe and GdGaGe follow the modified Curie-Weiss law, $\chi$(T) = $C/(T - \theta_P)$+ $\chi_0$, where $C$ is the Curie constant, $\theta_P$ is the paramagnetic Curie temperature, $\chi_0$ is temperature-independent magnetic susceptibility. The inverse susceptibilities 1/$\chi$(T) of NdGaGe and GdGaGe can be fitted using the modified Curie-Weiss law above 50 K (Fig. S2), and the fitting parameters are summarized in Table S2. The effective magnetic moment $\mu_{eff}$ determined by the Curie constant ( = $(3k_BC/N_A)^{1/2}$) for NdGaGe closely approximates the theoretical value (3.62 $\mu_B$ for free $Nd^{3+}$ ion), and is slightly larger than that for free $Gd^{3+}$ ion ( ~ 7.94 $\mu_B$) in GdGaGe. For NdGaGe, the negative $\theta_P$ (= -2.39 K for $B // a$) and positive $\theta_p$ (= 16.04 K for $B // c$) indicates the antiferromagnetic interactions between the Nd moments in plane and ferromagnetic interactions along out-of-plane direction, respectively. Furthermore, in GdGaGe, the negative Weiss temperatures ($\theta_P$ = -41.98 K for $B // a$ and $\theta_p$ = -42.91 K for $B // c$) reveal the presence of antiferromagnetic interactions for both crystallographic axes.

The isothermal magnetization M($B$) curves of NdGaGe and GdGaGe single crystal along $B // c$ across the temperature range 2 – 50 K are shown in Fig. 2(b) and 2(d), respectively. At 2 K, the M($B$) curve of NdGaGe shows rapid saturation behavior at extremely low magnetic field (2.29 T) and magnetic hysteresis with a small coercive field (≈ 0.025 T). As the temperature increasing, the saturation magnetization gradually decreases, and the M($B$) curves show linear behavior above $T_N$. Furthermore, when $B // a$, the M($B$) curve at 2 K exhibits almost linear behavior in the high-field region without saturation up to 7 T (Fig. S3(a)). As demonstrated in Fig. 2(d), the M($B$) curve of GdGaGe demonstrates a sharp upturn with increasing magnetic field, revealing a spin-flop transition at $B_{flop}$ ≈ 6.1 T at 2 K (inset of Fig. 2(d)). Beyond $B_{flop}$, the magnetization shows a monotonic increase without reaching saturation up to 7 T. With increasing temperature, $B_{flop}$ slightly shifts toward higher magnetic field and eventually vanishes above $T_N$, where M($B$) exhibits a linear field dependence characteristic in the paramagnetic state. When $B // a$, M($B$) shows nearly linear increase and no saturation up to 7 T at 2 K (Fig. S2(b)).

Based on the intriguing magnetic behaviors for $B // c$ compared with those for $B // a$, we mainly focus on the magnetotransport properties of NdGaGe and GdGaGe for $B // c$. The two compounds reveal distinct field-dependent behaviors. The magnetoresistance MR (= [$\rho_{xx}(B)$ - $\rho_{xx}(0)$]/$\rho_{xx}(0)$ × 100%) of NdGaGe with the magnetic field along the [001] direction is presented in Fig. 3(a). At $T$ = 2 K, NdGaGe enters into an AFM state and exhibits a significant negative MR behavior, attributed to the suppression of spin-disorder scattering as the magnetic moments become aligned under the applied field. As the magnetic field increases to approximately 1.7 T, the spin magnetic moments become fully polarized along the magnetic field direction, resulting in the saturation of MR. With increasing temperature, the negative MR increases monotonically, reaching a maximum value of approximately 13% at 7 K. Above $T_N$, the MR gradually decreases with increasing temperature. The Hall resistivity $\rho_{yx}(B)$ as a function of magnetic field under various temperatures is shown in Fig. 3(b). The $\rho_{yx}(B)$ initially exhibits a rapid step-like increase under low magnetic field, followed by a linear field dependence behavior at high field. As the temperature increases, $\rho_{yx}(B)$ curves gradually bend and eventually become linear in magnetic field. The similarity between $\rho_{yx}(B)$ and M(B) curves suggests the presence of an AHE in NdGaGe, which will be the focus of subsequent discussion.

As shown in Fig. 3(c), the MR of GdGaGe below $T_N$ is initially positive under low magnetic fields due to the effect of the Lorenz force, followed by an abrupt decrease around the field of $B_{flop}$ and then becomes negative. The negative MR behavior is attributed to the reduction in spin scattering after spin-flop transition. As shown in Fig. 3(d), $\rho_{yx}(B)$ exhibits a linear relationship with the magnetic field at 2 K. However, this linearity deviates near the spin-flop transition field and does not saturate until 7 T. The anomaly around $B_{flop}$ disappears above $T_N$, consistent with those observed in M(B) and MR curves.

In general, the total Hall resistivity is generally composed of two components:

$$\rho_{yx} = \rho_{yx}^o + \rho_{yx}^A = R_0 B + S_H \rho_{xx}^2 M \tag{1}$$

where the first term $\rho_{yx}^o$ ($R_0 B$) is the normal Hall resistivity and the second term $\rho_{yx}^A$ ($S_H \rho_{xx}^2 M$) is the anomalous Hall resistivity. $R_0$ and $S_H \rho_{xx}^2$ represents the ordinary and anomalous Hall coefficient, respectively.

For NdGaGe, we analyzed the relationship between $\rho_{yx}/B$ and $\rho_{xx}^2 M/B$ across different temperatures (Fig. 4(a)), obtaining $R_0$ from the intercept and the scaling factor $S_H$ from the slope according to the linear fitting of scaling curves at high field. The observed linear behavior confirms that the AHE in NdGaGe originates predominantly from the intrinsic mechanism. Furthermore, the $R_0$ above 10 K can be directly determined from the slope of $\rho_{yx}(B)$ curves. As shown in Fig. 4(b), the positive sign of $R_0$ (T) in all temperatures indicates the hole-type carriers are dominate in NdGaGe. Correspondingly, the carrier density $n$ calculated using the single-band model $R_0 = 1/en$ is in the order of $10^{22}$ cm$^{-3}$. Moreover, in the inset of Fig. 4(a), the $S_H$ remains approximately constant with 0.075 ~ 0.081 V$^{-1}$, further suggesting that the intrinsic mechanism is dominant in AHE for NdGaGe [2,56,57].

The AHC ($\sigma_{xy}^A$) can be evaluated by the relation $\sigma_{xy}^A = \rho_{yx}^A / \left[ \left(\rho_{yx}^A\right)^2 + \rho_{xx}^2 \right]$. $\sigma_{xy}^A$ remains almost unchanged in the temperature range of 2 – 10 K and reaches a maximum value of approximately 368 Ω$^{-1}$cm$^{-1}$ at 5 K, as demonstrated in Fig. 4(c). The anomalous Hall angle ($\sigma_{xy}^A / \sigma_{xx}$) is also insensitive to temperature and reaches ~ 0.9 % at 2 K (the inset of Fig. 4(c)). Additionally, $\sigma_{xy}^A$ exhibits a weak dependence on $\sigma_{xx}$, with a scaling exponent of 0.04. Therefore, the large AHC in NdGaGe primarily originates from the intrinsic Berry-phase contribution.

However, for GdGaGe, analyzing the $\rho_{yx}^A$ through the scaling relationship of $\rho_{yx}/B$ vs. $\rho_{xx}^2 M/B$ is challenging in the high-field region. Instead, a linear fit was performed above $B_{flop}$ according to Eq. 1, where the slope and zero-field intercept yield $R_0$ and $\rho_{yx}^A$, respectively. The positive $R_0$ demonstrates that the dominant carriers are hole-type in GdGaGe. The estimated $\rho_{yx}^A$ is weak in the whole temperature range, reaching about 0.027 μΩ cm at 2 K. The corresponding $\sigma_{xy}^A$ is approximately 23 Ω$^{-1}$ cm$^{-1}$.

Rare-earth ternary intermetallics exhibit complex relationships between their R composition and magnetic structure, displaying magnetic behaviors ranging from simple paramagnetism to complex magnetic phase transitions, which are primarily governed by the hybridization strength between 4$f$ rare-earth irons and conduction electrons. In the family of materials adopting non-centrosymmetric LaPtSi-type structures, antiferromagnets NdGaGe and GdGaGe display distinct magnetic and electrical transport properties, suggesting that the rare-earth ions significantly influence the magnetic structure and electronic band structure. The AHE observed in NdGaGe and GdGaGe show significant quantitative differences, attributable to the modulation effects of different rare-earth elements in the RGaGe family.

In summary, we comprehensively investigated the high-quality NdGaGe and GdGaGe single crystals. Both compounds show antiferromagnetic phase transition ($T_N$ ~ 7.6 K for NdGaGe, $T_N$ ~ 22.4 K for GdGaGe). NdGaGe exhibits strong magnetic anisotropy, while GdGaGe shows weak isotropic behavior with a spin-flop transition under magnetic field. Hall measurements reveal hole-dominated conduction in both compounds. NdGaGe shows a large AHC (~ 368 $\Omega^{-1}$ cm$^{-1}$), which is dominated by the intrinsic mechanism. The results in GdGaGe and NdGaGe highlight the crucial role of rare-earth elements in determining the electronic structure and associated properties of the RGaGe family. Our systematic investigation deepens the understanding of the correlation between electronic structure and magnetism in these materials, providing a candidate platform for next-generation spintronic device applications.

See the supplementary material for detailed magnetic measurement and structure parameter of NdGaGe and GdGaGe.

This work was supported by the National Natural Science Foundation of China (Grant Nos. 52272265 and 12404161), the National Key R&D Program of China (Grant No. 2023YFA1607400) and the Shanghai Sailing Program (23YF1426800). The authors thank the support from Analytical Instrumentation Center (# SPSTAIC10112914), SPST, ShanghaiTech University.

## REFERENCES

[1] N. Nagaosa, J. Sinova, S. Onoda, A. H. MacDonald, and N. P. Ong, Rev. Mod. Phys. **82**, 1539 (2010).
[2] T. Jungwirth, Q. Niu, and A. MacDonald, Phys. Rev. Lett. **88**, 207208 (2002).
[3] A. M. Ionescu and H. Riel, nature **479**, 329 (2011).
[4] I. Žutić, J. Fabian, and S. D. Sarma, Rev. Mod. Phys. **76**, 323 (2004).
[5] D. D. Awschalom and M. E. Flatté, Nature physics **3**, 153 (2007).

[6] R. Karplus and J. Luttinger, Phys. Rev. **95**, 1154 (1954).

[7] M. Onoda and N. Nagaosa, Phys. Rev. Lett. **90**, 206601 (2003).

[8] D. Xiao, M.-C. Chang, and Q. Niu, Rev. Mod. Phys. **82**, 1959 (2010).

[9] Q. Wang, S. Sun, X. Zhang, F. Pang, and H. Lei, Phys. Rev. B **94**, 075135 (2016).

[10] Q. Wang, Y. Xu, R. Lou, Z. Liu, M. Li, Y. Huang, D. Shen, H. Weng, S. Wang, and H. Lei, Nat. Commun. **9**, 1 (2018).

[11] S. N. Guin, K. Manna, J. Noky, S. J. Watzman, C. Fu, N. Kumar, W. Schnelle, C. Shekhar, Y. Sun, and J. Gooth, NPG Asia Materials **11**, 16 (2019).

[12] L. Ye, M. Kang, J. Liu, F. Von Cube, C. R. Wicker, T. Suzuki, C. Jozwiak, A. Bostwick, E. Rotenberg, and D. C. Bell, Nature **555**, 638 (2018).

[13] E. Liu, Y. Sun, N. Kumar, L. Muechler, A. Sun, L. Jiao, S.-Y. Yang, D. Liu, A. Liang, and Q. Xu, Nature physics **14**, 1125 (2018).

[14] S. Nakatsuji, N. Kiyohara, and T. Higo, Nature **527**, 212 (2015).

[15] A. Markou, D. Kriegner, J. Gayles, L. Zhang, Y.-C. Chen, B. Ernst, Y.-H. Lai, W. Schnelle, Y.-H. Chu, and Y. Sun, Phys. Rev. B **100**, 054422 (2019).

[16] J. Smit, Physica **21**, 877 (1955).

[17] J. Smit, Physica **24**, 39 (1958).

[18] L. Berger, Phys. Rev. B **2**, 4559 (1970).

[19] S.-Y. Xu, N. Alidoust, G. Chang, H. Lu, B. Singh, I. Belopolski, D. S. Sanchez, X. Zhang, G. Bian, and H. Zheng, Sci. Adv. **3**, e1603266 (2017).

[20] W. Cao, N. Zhao, C. Pei, Q. Wang, Q. Zhang, T. Ying, Y. Zhao, L. Gao, C. Li, and N. Yu, Phys. Rev. B **105**, 174502 (2022).

[21] G. Chang, B. Singh, S.-Y. Xu, G. Bian, S.-M. Huang, C.-H. Hsu, I. Belopolski, N. Alidoust, D. S. Sanchez, and H. Zheng, Phys. Rev. B **97**, 041104 (2018).

[22] H.-Y. Yang, B. Singh, J. Gaudet, B. Lu, C.-Y. Huang, W.-C. Chiu, S.-M. Huang, B. Wang, F. Bahrami, and B. Xu, Phys. Rev. B **103**, 115143 (2021).

[23] H. Hodovanets, C. Eckberg, P. Zavalij, H. Kim, W.-C. Lin, M. Zic, D. Campbell, J. Higgins, and J. Paglione, Phys. Rev. B **98**, 245132 (2018).

[24] T. Suzuki, L. Savary, J.-P. Liu, J. W. Lynn, L. Balents, and J. G. Checkelsky, Science **365**, 377 (2019).

[25] M. S. Alam, A. Fakhredine, M. Ahmad, P. Tanwar, H.-Y. Yang, F. Tafti, G. Cuono, R. Islam, B. Singh, and A. Lynnyk, Phys. Rev. B **107**, 085102 (2023).

[26] P. Puphal, V. Pomjakushin, N. Kanazawa, V. Ukleev, D. J. Gawryluk, J. Ma, M. Naamneh, N. C. Plumb, L. Keller, and R. Cubitt, Phys. Rev. Lett. **124**, 017202 (2020).

[27] L. Wu, S. Chi, H. Zuo, G. Xu, L. Zhao, Y. Luo, and Z. Zhu, npj Quantum Mater. **8**, 4 (2023).

[28] H.-Y. Yang, B. Singh, B. Lu, C.-Y. Huang, F. Bahrami, W.-C. Chiu, D. Graf, S.-M. Huang, B. Wang, and H. Lin, APL Mater. **8** (2020).

[29] M. Lyu, J. Xiang, Z. Mi, H. Zhao, Z. Wang, E. Liu, G. Chen, Z. Ren, G. Li, and P. Sun, Phys. Rev. B **102**, 085143 (2020).

[30] D. S. Sanchez, G. Chang, I. Belopolski, H. Lu, J.-X. Yin, N. Alidoust, X. Xu, T. A. Cochran, X. Zhang, and Y. Bian, Nat. Commun. **11**, 3356 (2020).

[31] B. Meng, H. Wu, Y. Qiu, C. Wang, Y. Liu, Z. Xia, S. Yuan, H. Chang, and Z. Tian, APL Mater. **7** (2019).

[32] M. Lyu, Z. Wang, K. Ramesh Kumar, H. Zhao, J. Xiang, and P. Sun, J. Appl. Phys. **127** (2020).

[33] J. Gaudet, H.-Y. Yang, S. Baidya, B. Lu, G. Xu, Y. Zhao, J. A. Rodriguez-Rivera, C. M. Hoffmann, D. E. Graf, and D. H. Torchinsky, Nat. Mater. **20**, 1650 (2021).

[34] N. Zhang, X. Ding, F. Zhan, H. Li, H. Li, K. Tang, Y. Qian, S. Pan, X. Xiao, and J. Zhang, Physical Review Research **5**, L022013 (2023).

[35] J.-F. Wang, Q.-X. Dong, Z.-P. Guo, M. Lv, Y.-F. Huang, J.-S. Xiang, Z.-A. Ren, Z.-J. Wang, P.-J. Sun, and G. Li, Phys. Rev. B **105**, 144435 (2022).

[36] J.-F. Wang, Q.-X. Dong, Y.-F. Huang, Z.-S. Wang, Z.-P. Guo, Z.-J. Wang, Z.-A. Ren, G. Li, P.-J. Sun, and X. Dai, Phys. Rev. B **108**, 024423 (2023).

[37] C. Li, J. Zhang, Y. Wang, H. Liu, Q. Guo, E. Rienks, W. Chen, F. Bertran, H. Yang, and D. Phuyal, Nat. Commun. **14**, 7185 (2023).

[38] C. Dhital, R. L. Dally, R. Ruvalcaba, R. Gonzalez-Hernandez, J. Guerrero-Sanchez, H. B. Cao, Q. Zhang, W. Tian, Y. Wu, and M. D. Frontzek, Phys. Rev. B **107**, 224414 (2023).

[39] J. Zhao, W. Liu, A. Rahman, F. Meng, L. Ling, C. Xi, W. Tong, Y. Bai, Z. Tian, and Y. Zhong, New J. Phys. **24**, 013010 (2022).

[40] H.-Y. Yang, J. Gaudet, R. Verma, S. Baidya, F. Bahrami, X. Yao, C.-Y. Huang, L. DeBeer-Schmitt, A. A. Aczel, and G. Xu, Phys. Rev. Mater. **7**, 034202 (2023).

[41] W. Cao, Y. Su, Q. Wang, C. Pei, L. Gao, Y. Zhao, C. Li, N. Yu, J. Wang, and Z. Liu, Chin. Phys. Lett. **39**, 047501 (2022).

[42] L. Xu, H. Niu, Y. Bai, H. Zhu, S. Yuan, X. He, Y. Han, L. Zhao, Y. Yang, and Z. Xia, J. Phys.: Condens. Matter **34**, 485701 (2022).

[43] X. Yao, J. Gaudet, R. Verma, D. E. Graf, H.-Y. Yang, F. Bahrami, R. Zhang, A. A. Aczel, S. Subedi, and D. H. Torchinsky, Phys. Rev. X **13**, 011035 (2023).

[44] Y. Gao, S. Lei, E. M. Clements, Y. Zhang, X.-J. Gao, S. Chi, K. T. Law, M. Yi, J. W. Lynn, and E. Morosan, arXiv preprint arXiv:2310.09364 (2023).

[45] E. Cheng, L. Yan, X. Shi, R. Lou, A. Fedorov, M. Behnami, J. Yuan, P. Yang, B. Wang, and J.-G. Cheng, Nat. Commun. **15**, 1467 (2024).

[46] A. P. Sakhya, C.-Y. Huang, G. Dhakal, X.-J. Gao, S. Regmi, B. Wang, W. Wen, R.-H. He, X. Yao, and R. Smith, Phys. Rev. Mater. **7**, L051202 (2023).

[47] J. Kunze, M. Köpf, W. Cao, and Y. Qi, Phys. Rev. B **109**, 195130 (2024).

[48] K. Wang, W. Shi, W. Cao, X. Yang, Z. Lv, C. Peng, C. Chen, D. Liu, H. Yang, and L. Yang, arXiv preprint arXiv:2503.12887 (2025).

[49] D. Ram, L. Joshi, and Z. Hossain, Journal of Magnetism and Magnetic Materials **605**, 172326 (2024).

[50] J. Gong, H. Wang, K. Han, X.-Y. Zeng, X.-P. Ma, Y.-T. Wang, J.-F. Lin, X.-Y. Wang, and T.-L. Xia, Phys. Rev. B **109**, 024434 (2024).

[51] A. Sarkis and E. Callen, Phys. Rev. B **26**, 3870 (1982).

[52] R. Skomski and D. Sellmyer, Journal of Rare Earths **27**, 675 (2009).

[53] C. E. Patrick, S. Kumar, G. Balakrishnan, R. S. Edwards, M. R. Lees, L. Petit, and J. B. Staunton, Phys. Rev. Lett. **120**, 097202 (2018).

[54] D. Ram, J. Singh, M. Hooda, K. Singh, V. Kanchana, D. Kaczorowski, and Z. Hossain, Phys. Rev. B **108**, 235107 (2023).

[55] T. Kurumaji, M. Gen, S. Kitou, and T.-h. Arima, Phys. Rev. B **110**, 064409 (2024).

[56] P. Dheer, Phys. Rev. **156**, 637 (1967).

[57] J.-P. Jan and H. Gijsman, Physica **18**, 339 (1952).

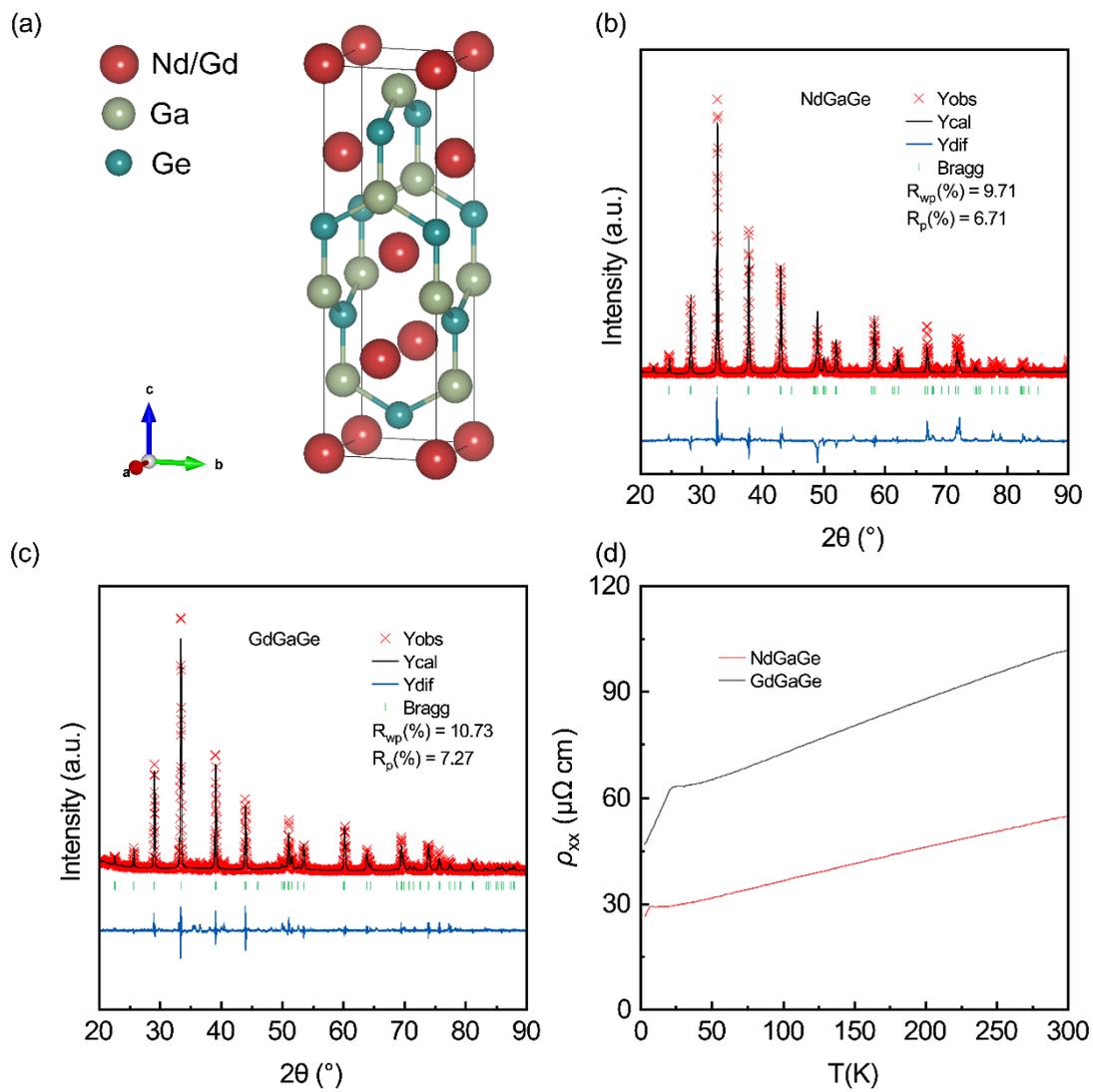

Fig. 1. (a) The crystal structure of NdGaGe and GdGaGe. (b) and (c) Rietveld refinement of the powder XRD patterns of NdGaGe and GdGaGe, respectively. (d) Temperature dependence of $\rho_{xx}(T)$ in the temperature range from 1.8 K to 300 K.

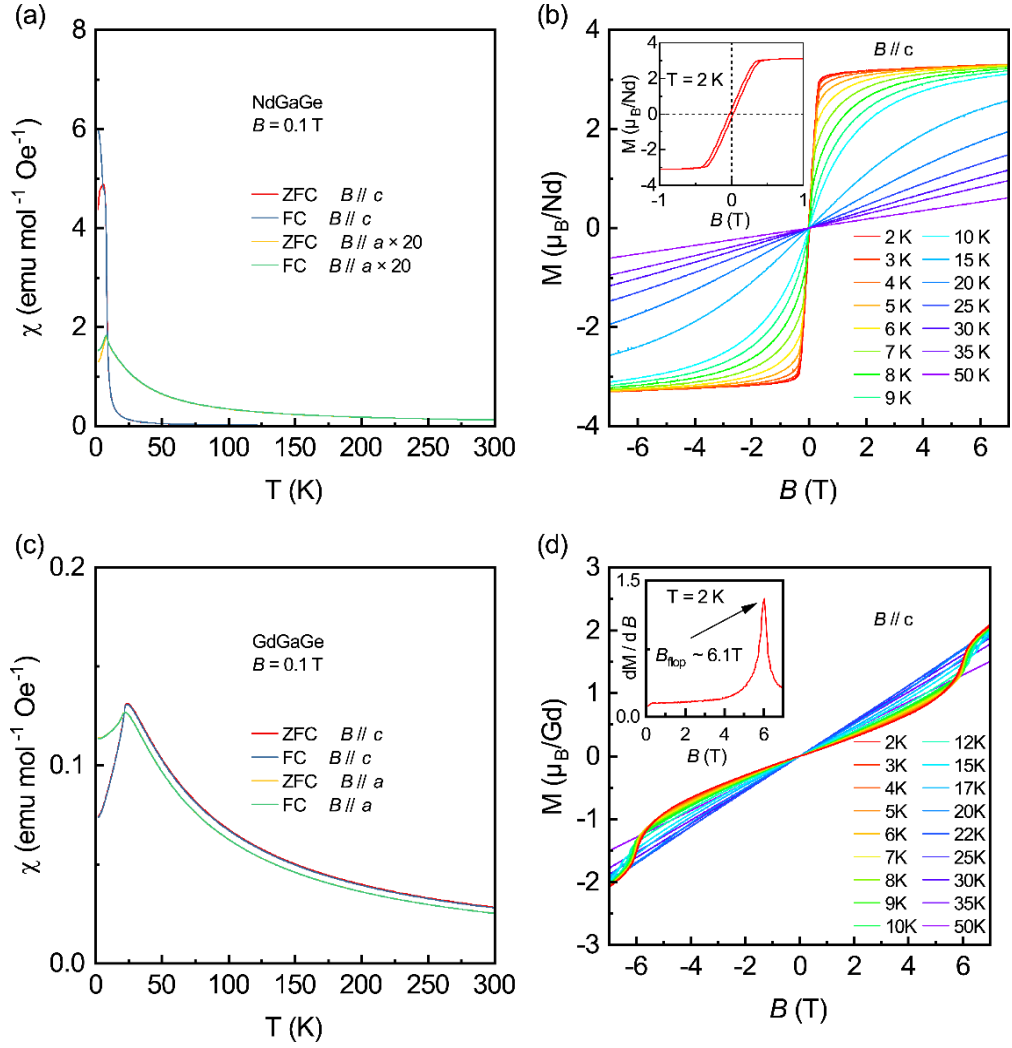

Fig. 2. (a) and (c) Temperature dependence of the magnetic susceptibility χ(T) of NdGaGe and GdGaGe with ZFC and FC modes (*B* = 0.1 T) for *B* // *c* and *B* // *a*. (b) Magnetic field dependent magnetization M(*B*) of NdGaGe at various temperatures for *B* // *c*. Inset of (b): magnetic hysteresis loop at 2 K with *B* // *c* (d) Magnetic field dependent magnetization M(*B*) of GdGaGe at various temperatures for *B* // *c*. Inset of (d): the derivative of M(*B*) at 2 K, the peak at $B_{flop}$ ≈ 6.1 T corresponds to the spin-flop transition.

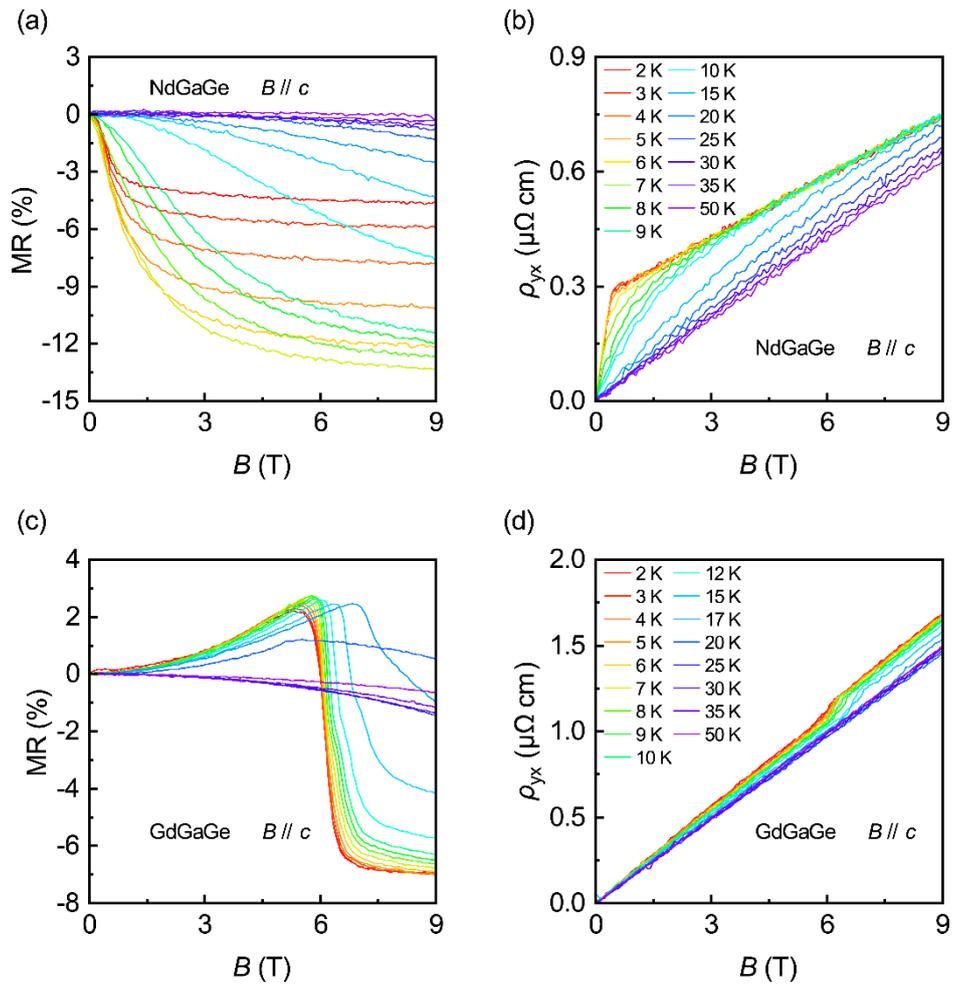

Fig. 3. (a) Magnetoresistance (MR) and (b) Hall resistivity as a function of magnetic field at various temperatures for NdGaGe. The magnetoresistance is defined as MR = [$\rho(B) - \rho(0)$]/ $\rho(0)$ × 100% in which $\rho(B)$ and $\rho(0)$ represent the resistivity with and without $B$, respectively. (c) and (d) The MR and Hall resistivity for GdGaGe, respectively.

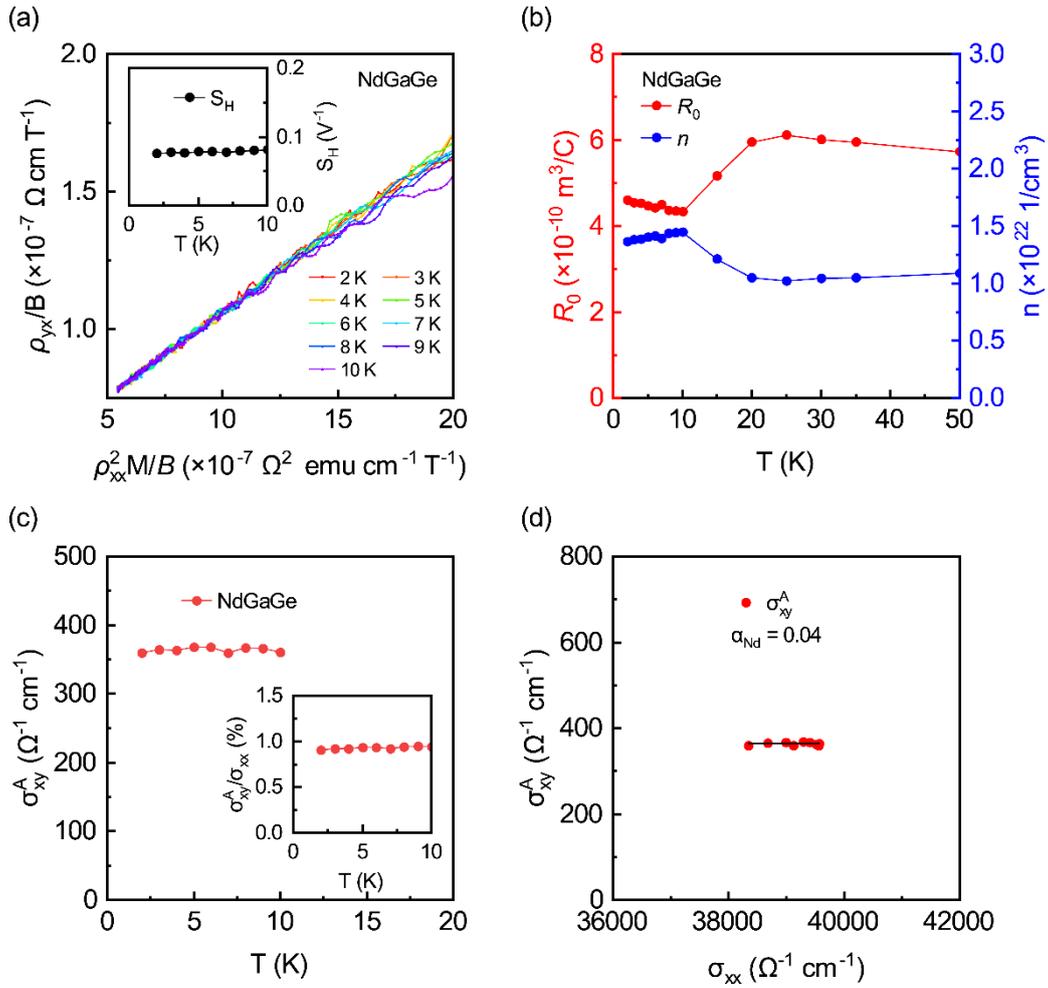

Fig. 4. (a) The scaling behavior of $\rho_{yx}/B$ versus $\rho_{xx}^2 M/B$ at different temperatures for NdGaGe. Inset of (a): Temperature dependence of $S_H(T)$ for NdGaGe. (b) Temperature dependence of ordinary Hall coefficient $R_0$ and carrier density $n$ for NdGaGe. (c) Temperature dependence of AHC for NdGaGe. Inset of (c): The temperature dependence of anomalous Hall angle (AHA) $\sigma_{xy}^A/\sigma_{xx}$. (d) Plot of $\sigma_{yx}^A$ against $\sigma_{xx}$ for NdGaGe.